\documentclass[a4paper,11pt]{article}
\usepackage{pos}

\newcommand{\lhc}{{\rm lhc}}

\newcommand{\be}{\begin{equation}} 
\newcommand{\ee}{\end{equation}}
\newcommand{\bea}{\begin{eqnarray}} 
\newcommand{\eea}{\end{eqnarray}}

\newcommand{\nno}{\nonumber\\}

\newcommand{\im}{\operatorname{Im}}

\definecolor{green}{rgb}{0,0.6,0}

\newcommand{\kk}{k^{\prime 2}}
\newcommand{\mex}{m_\text{ex}^2}
\newcommand{\al}{&\!\!\!\!}
\newcommand{\intlhc}{\int_{-\infty}^{k_\text{lhc}^2}}

\title{Effective range expansion with the left-hand cut and its application to the $T_{cc}(3875)$}

\author*[a]{Meng-Lin Du}
\author[b,c,d]{Feng-Kun Guo}
\author[a]{Bing Wu}

\affiliation[a]{School of Physics, University of Electronic Science and Technology of China,\\
 Chengdu 611731, China}

\affiliation[b]{CAS Key Laboratory of Theoretical Physics, Institute of Theoretical Physics, \\Chinese Academy of Sciences, Beijing 100190, China}

\affiliation[c]{School of Physical Sciences, University of Chinese Academy of Sciences,\\ Beijing 100049, China}

\affiliation[d]{Peng Huanwu Collaborative Center for Research and Education,\\ Beihang University, Beijing 100191, China}

\emailAdd{du.ml@uestc.edu.cn}
\emailAdd{fkguo@itp.ac.cn}
\emailAdd{wu.bing@uestc.edu.cn}

\abstract{The validity range of the widely used traditional effective range expansion can be severely limited by the presence of a left-hand cut near the two-particle threshold. Such a left-hand cut emerges in two-particle scattering processes involving either a light particle exchange in the $t$-channel or a particle exchange with a mass slightly heavier than the mass difference of the two particles in the $u$-channel, which occurs in a wide range of physical systems. 
We propose a new parameterization for the low-energy scattering amplitude that incorporates these left-hand cuts arising from particle exchange diagrams. This parameterization extends the convergence radius of the effective range expansion beyond the branch point of the left-hand cut and is applicable to a broad range of systems. 
The parameterization enables the extraction of coupling strengths between the exchange particle and the scattering particles, and reveals amplitude zeros resulting from the interplay between short- and long-range interactions. We demonstrate the effectiveness of this new parameterization through its application to $DD^*$ scattering with meson masses obtained in a lattice QCD calculation.}

\FullConference{The 11th International Workshop on Chiral Dynamics (CD2024)\\
 26-30 August 2024\\
Ruhr University Bochum, Germany\\}

\begin{document}
\maketitle

\section{Introduction}\label{sec:intro}

The effective range expansion (ERE)~\cite{Bethe:1949yr,Blatt:1949zz} is a very general parameterization of low-energy amplitudes in quantum mechanics that is extensively utilized across different fields of physics, including particle physics, nuclear physics, and cold atom physics. It provides a simple and efficient way to extract dynamical information about the interaction between two particles in the low-energy regime. ERE is a direct consequence of the unitarity and analyticity of the amplitude in the vicinity of the two-particle threshold.

Unitarity requires that the imaginary part of the inverse of the two-particle scattering amplitude $f$ in the elastic region is proportional to the magnitude of the relative momentum $k$:
\bea\label{eq:imf}
\text{Im}\frac{1}{f}=-k,
\eea
in nonrelativistic kinematics. The imaginary part $k=\sqrt{2\mu E}$ leads to a cut in the complex energy plane of the amplitude, known as the unitarity cut or right-hand cut (rhc). Here, $\mu$ is the reduced mass of the two scattering particles, and $E$ is the center-of-mass energy of the system relative to the threshold. Analyticity of the amplitude implies that in the vicinity of the threshold, the inverse of the amplitude is analytic except for the unitarity cut and thus can be expanded in powers of $k^2=2\mu E$. For an $S$-wave amplitude, ERE has the form:
\bea\label{eq:ere}
\frac{1}{f} = k\cot \delta -ik= \frac{1}{a}+\frac{1}{2}r k^2-ik+\mathcal{O}(k^4),
\eea
where $\delta$ is the scattering phase shift, and the parameters $a$ and $r$ are the well-known scattering length and effective range, respectively. The expansion~\eqref{eq:ere} is limited by the presence of other singularities, such as three-body threshold or left-hand cut (lhc) branch points. The lhc generated by particle exchange is particularly important in two cases: when a light particle is exchanged in the $t$-channel, or when a particle slightly heavier than the mass difference of the two scattering particles is exchanged in the $u$-channel. In these cases, the branch point can be located very close to the threshold, severely restricting the validity range of the expansion~\eqref{eq:ere}.

ERE is particularly widely used in hadronic and nuclear physics due to the nonperturbative nature of the strong interaction at low energies, where unitarity plays a crucial role in the resonance region. In the past two decades, numerous resonances have been experimentally observed in hadron physics~\cite{Guo:2017jvc,Chen:2022asf,ParticleDataGroup:2024cfk}. 
Many of these resonances are close to two-hadron thresholds, making their corresponding two-hadron amplitudes potentially suitable for analysis using ERE. 
However, the one-particle-exchange (OPE), particularly the one-pion-exchange, potential is commonly encountered in hadron calculations, such as in $NN$, $BB^*$, $B\bar{B}^*$, $ND^*$, $N\bar{D}^*$, $\Sigma_{(c)}D^{*}$, $\Sigma_c\bar{D}^*$, and others. Notably, the unphysical pion mass used in LQCD simulations is usually larger than the physical value. 
This leads to a stable $D^*$ and a left-hand cut caused by the one-pion-exchange in the $DD^*$ and $D\bar{D}^*$ scattering amplitudes in LQCD calculations, e.g.~\cite{Prelovsek:2013cra,Prelovsek:2014swa,CLQCD:2019npr,Padmanath:2022cvl,Chen:2022vpo,Lyu:2023xro,Li:2024pfg,Sadl:2024dbd}. 
The left-hand cut caused by OPE near the two-hadron threshold invalidates the applicability of ERE~\eqref{eq:ere} in this energy region.

In lattice simulations, only discrete energy levels are obtained due to the finite volume. The L\"uscher formalism~\cite{Luscher:1990ux,Luscher:1985dn,Luscher:1986pf} provides a bridge to extract the scattering phase shift, and consequently the ERE parameters, from the energy levels in finite volumes. The original L\"uscher formalism does not account for the lhc, and therefore the extracted phase shifts can only be real values.
However, in certain LQCD calculations, the finite-volume energy levels have been found to be lower than the corresponding infinite-volume scattering lhc branch point, for example in~\cite{Green:2021qol} for baryon-baryon scattering and in~\cite{Padmanath:2022cvl,Lyu:2023xro} for $DD^*$ scattering.
The presence of the lhc invalidates both the original L\"uscher formalism and ERE used in LQCD analyses. It has been demonstrated in~\cite{Du:2023hlu} that the presence of the lhc can induce significant effects on the phase shift and pole structures.
Note that different strategies have been proposed to rescue the L\"uscher formalism \cite{Meng:2023bmz, Raposo:2023oru,Hansen:2024ffk,Bubna:2024izx}.

\section{The left-hand cut}\label{sec:lhc}

\begin{figure}[htb!]
\centering
\includegraphics[width=0.6\linewidth]{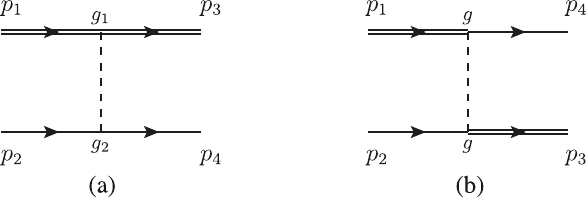}
\caption{Two particles interacting through a $t$-channel (left) and a $u$-channel (right) particle exchange, where $g_{(i)}$ denote the corresponding coupling constants.}
\label{fig:feyndiag}
\end{figure}

The lhc is a general feature of partial wave scattering amplitudes, originating from crossed-channel singularities. It is important to note that this cut exists in the complex energy plane of the scattering amplitude in infinite volume, where the permitted energy is continuous above the two-particle threshold.
In finite volume, however, two-particle scattering is not well-defined, and the permitted energy levels of the system are discrete.
The lhc arising from exchanging one nearly on-shell particle is especially significant since its branch point can be very close to the two-particle threshold, potentially having a substantial impact on physical quantities, as shown in~\cite{Du:2023hlu}.
In this talk, we focus specifically on this type of lhc, where the OPE potential corresponds to a long-range but still finite-range interaction. For elastic two-hadron scattering, there are two possible types of Feynman diagrams, corresponding to t- and u-channel exchanges, as shown in Fig.~\ref{fig:feyndiag}.

The lhc caused by the OPE diagram in Fig.~\ref{fig:feyndiag} exhibits a discontinuity of the partial wave potential across the real axis below the lhc branch point. For simplicity, we consider the case where the coupling between the scattering and exchanging particles is in $S$-wave with coupling constants $g_{(i)}$, as shown in Fig.~\ref{fig:feyndiag}. 
For an $S$-wave scattering between the two particles, the corresponding partial wave is given by
\newcommand{\td}{\text{d}}
\bea
g_1g_2 L_t(s) \al=\al \frac{g_1g_2}{2}\int\frac{1}{t-\mex}\td\cos\theta = -\frac{g_1g_2 s}{\lambda(s,m_1,m_2)}\log\left(\frac{s-2(m_1^2+m_2^2)+\mex+\frac{(m_1^2-m_2^2)^2}{s}}{\mex}\right),\nno
g^2L_u(s)\al=\al\frac{g^2}{2}\int\frac{1}{u-\mex}\td\cos\theta=-\frac{g^2s}{\lambda(s,m_1^2,m_2^2)}\log\left(\frac{s+\mex-2(m_1+m_2)^2}{\mex-(m_1^2-m_2^2)^2/s}\right),
\eea
for the $t$- and $u$-channel, respectively. Here $s=(m_1+m_2+E)^2$, $t=(p_1-p_3)^2$, $u=(p_1-p_4)^2$, and $\lambda(a,b,c)=a^2+b^2+c^2-2ab-2bc-2ca$ is the K\"all\'en function, with $m_1$ and $m_2$ representing the masses of the scattering particles, and $m_\text{ex}$ the mass of the exchanging particle.

In nonrelativistic kinematics, the corresponding potential can be simplified to \cite{Du:2024snq}
\bea\label{eq:L}
L_t(k^2) = -\frac{1}{4k^2}\log\frac{\mex/4+k^2}{\mex/4}, \quad
L_u(k^2) 
\approx  -\frac{1}{4k^2}\log\frac{\mu_+^2/4+k^2}{\mu_+^2/4+ \eta^2 k^2},
\eea
where $\eta = |m_1-m_2|/(m_1+m_2)$ and $\mu_+^2=4m_1m_2\left(\mex-(m_1-m_2)^2\right)/(m_1+m_2)^2$. The expression for $L_u(k^2)$ reduces to that of $L_t(k^2)$ in the limit $\Delta=m_1-m_2\to 0$. From Eq.~\eqref{eq:L}, one can deduce that the nearest lhc branch point is located at $k_\text{lhc}^2=-\mex/4$ and $-\mu_+^2/4$ for the $t$- and $u$-channel exchange, respectively. 
For $u$-channel exchanges, the condition that the lhc branch point is located below the threshold requires $\mex>(m_1-m_2)^2$. Otherwise, the heavier particle can decay into the other particle and the exchanging particle, leading to a three-body cut, which is beyond the scope of this work. 

The form of the logarithm in Eq.~\eqref{eq:L} of the $S$-wave scattering remains unchanged for different partial waves of the coupling vertices. For $P$-wave couplings, the corresponding partial wave projection of the OPE has the form
\bea\label{eq:L:P}
L^P_t(k^2)\al = \al \frac{1}{2}\int\frac{(\vec{p}_1-\vec{p}_3)^2}{t-\mex}\td\cos\theta = \frac{1}{2}\int\frac{-t}{t-\mex}\td\cos\theta = -1-\mex L_t(k^2),\nno 
L^P_u(k^2)\al =\al \frac{1}{2}\int\frac{-u+(m_1^2-m_2^2)/(m_1+m_2+E)}{u-\mex}\td\cos\theta \approx -1-{\mu_\text{ex}^2}L_u(k^2),
\eea
with $\mu_\text{ex}^2=\mex-(m_1-m_2)^2$. Here the constant term ($-1$) has no cut and represents a short contact interaction. The lhc/discontinuity for the $P$-wave coupling is also given by the $L(k^2)$ function, as that for the $S$-wave vertices. Higher order couplings can be treated similarly, and for an $S$-wave scattering, the general lhc has the form of $L(k^2)$.
\begin{figure}[htb!]
\centering
\includegraphics[width=0.7\linewidth]{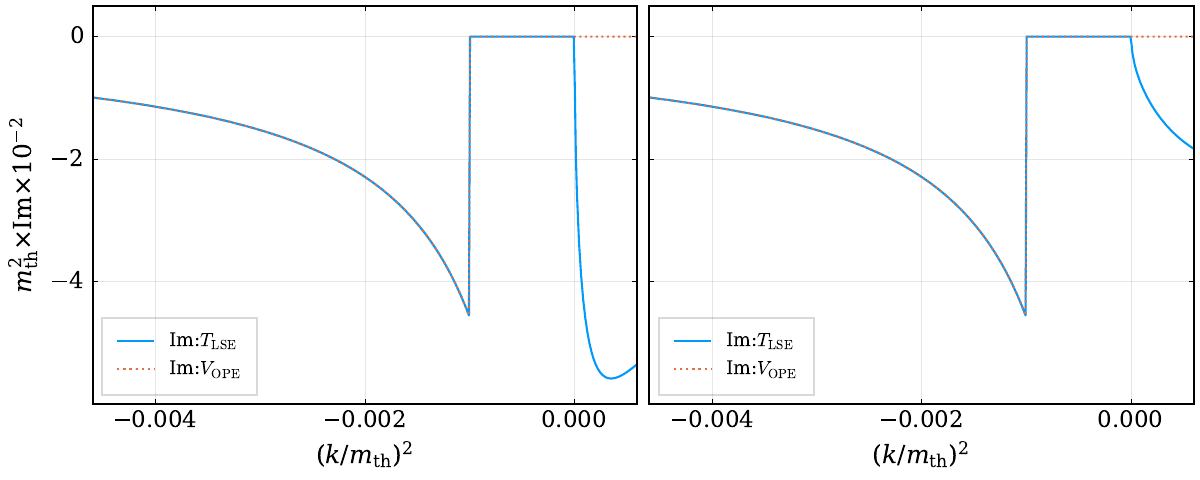}
\caption{The comparison between the imaginary part of the $T$-matrix amplitude from LSE and that of the tree-level OPE potential $V_\text{OPE}$. The left and right panels correspond to the left and right panel of Fig.~3 in~\cite{Du:2023hlu}, respectively.}
\label{fig:im:VT}
\end{figure}
The two-hadron scattering amplitude $T$ that satisfies unitarity can be obtained by solving the Lippmann-Schwinger equation (LSE)
\bea\label{eq:lse}
T(E,p,q) = V(E,p,k)+\int\frac{{\td}^3\vec{k}}{(2\pi)^3}V(E,p,q)\frac{1}{E-k^2/2\mu}T(E,k,q),
\eea
where the potential $V$ includes the lhc from the long-range OPE potential. Across the OPE lhc, the discontinuity of the $T$-matrix amplitude matches that of the tree-level potential $V$ used in the LSE. The second term on the right-hand side of Eq.~\eqref{eq:lse} does not contain the OPE lhc on the physical Riemann sheet. To illustrate this, Fig.~\ref{fig:im:VT} compares the imaginary parts of the $T$-matrix amplitude with the OPE potential $V_\text{OPE}$ from~\cite{Du:2023hlu}. Here we use the convention $f(k^2) = -\mu/(2\pi)T(E,k,k)$.

\section{The $N/D$ method and the ERE with the left-hand cut}\label{sec:nd}
\subsection{The $N/D$ method}

The scattering amplitude can be constructed from dispersive relations based on analyticity when the singularities of the amplitude are known. At low energies where only the unitarity cut and the lhc from OPE are present, we can construct the scattering amplitude satisfying unitarity and analyticity using the $N/D$ method~\cite{Chew:1960iv}. In this method, the partial wave amplitude $f$ is expressed as a ratio of two functions
\bea
f(k^2) = {n(k^2)}/{d(k^2)},
\eea
where the denominator $d(k^2)$ contains only the rhc, and $n(k^2)$ has only the lhc. From unitarity~\eqref{eq:imf}, we have the following dispersive relations
\bea\label{eq:dis:non}
\im d(k^2)\al = -k n(k^2),\qquad \qquad ~ \al k^2>0,\nno
\im n(k^2)\al =  d(k^2)\im f(k^2),\qquad \al k^2<k_\text{lhc}^2.
\eea
Both $n(k^2)$ and $d(k^2)$ can be simultaneously multiplied by any real analytical function without changing the amplitude $f(k^2)$ or the dispersive relations~\eqref{eq:dis:non}.

When the lhc is far from the threshold and can be neglected, $n(k^2)$ has no imaginary part along the real axis and can be set to $n(k^2)=1$. Since $\text{Im}d(k^2) = -k$, we have $d=-ik+r(k^2)$, where $r(k^2)$ is an analytical function. This function can be Taylor expanded in $k^2=2\mu E$ in the complex plane as $d(k^2)=-ik+r_0+r_1 k^2 +\mathcal{O}(k^4)$, which gives the well-known ERE.

In the presence of the lhc, the solution can be given by the dispersion relations 
\bea
n(k^2) \al = \al \bar{n}_m(k^2) +\frac{(k^2)^m}{\pi}\intlhc\frac{d(\kk)\im f(\kk)}{(\kk -k^2)(\kk)^m}\td \kk, \nno
d(k^2) \al = \al \bar{d}_n(k^2)-\frac{(k^2)^n}{\pi}\int_0^\infty \frac{k^\prime n(\kk)\td \kk}{(\kk-k^2)(\kk)^n},
\eea
where $\bar{n}_m(k^2)$ and $\bar{d}_n(k^2)$ are subtracted polynomials that ensure the interaction is well defined. Since $d(k^2)$ is real analytical below the threshold and $n(k^2)$ has an imaginary part along the lhc, a direct consequence of the $N/D$ method is that no virtual state can exist along the lhc. 
This is because the amplitude on the second Riemann sheet $f^\text{II}$ is defined as $f^\text{II}=\left(1/f^\text{I}+2ik\right)^{-1}=n/(d+2ikn)$, where the denominator $(d+2ikn)$ cannot be zero since $k$ is purely imaginary.

\subsection{The effective range expansion with the left-hand cut}

As mentioned above, along the OPE lhc, the imaginary part of the amplitude $f(k^2)$ equals that of the tree-level potential shown in Fig.~\ref{fig:feyndiag}, which means 
\bea 
\im f(k^2) = c\im L(k^2) = -\frac{c}{4k^2}\pi, \qquad\qquad k^2<k_{\lhc}^2,
\eea
where $c$ is a parameter quantifying the strength of the lhc. For the $S$-wave couplings, $c=g_1 g_2$ and $g^2$ for the $t$- and $u$-channel OPE in Fig.~\ref{fig:feyndiag}, respectively. For $P$-wave vertices in Eq.~\eqref{eq:L:P}, $c=-g_1g_2\mex$ and $-g^2\mu_\text{ex}^2$ for $t$- and $u$-channel, respectively.

In general, the integration of $n(k^2)$ and $d(k^2)$ rely on each other and the exact solution is inaccessible. However, in the low-energy region where only the lhc from the OPE is present, we can parameterize these two functions.
Since $d(k^2)$ is real analytic along the lhc, it can be parameterized as polynomials, denoted as $P(k^2)$. Then the dispersive integral can be carried out as \cite{Du:2024snq}
\bea
n(k^2) = \bar{n}^\prime (k^2)+P(k^2) L(k^2)=\bar{n}(k^2) + P(k^2)(L(k^2)-L_0),
\eea
where $\bar{n}^{(\prime)}(k^2)$ is a polynomial. The expression for $L(k^2)$ is given by $L_t(k^2)$ or $L_u(k^2)$ in Eq.~\eqref{eq:L}, depending on the exchange channel. Here $L_0=L(k^2=0)=-1/\mex$ (or $-/\mu_\text{ex}^2$) for $t$-channel (or $u$-channel) is subtracted to make the lhc term vanish at the threshold. Therefore,
\bea
d(k^2) \al = \al  \bar{d}^\prime(k^2)_m-ik \bar{n}(k^2)-\frac{(k^2)^m}{\pi}\int_0^\infty \frac{k^\prime L(\kk)P(\kk)}{(\kk-k^2)(\kk)^m}\td \kk \nno
\al = \al \bar{d}(k^2)-ik n(k^2) - P(k^2) d^R(k^2),
\eea
where the $d^R(k^2)$ function is given by 
\begin{align}
d^\text{R}_t(k^2) &= \frac{i}{4k}\log\frac{m_\text{ex}/2+ik}{m_\text{ex}/2-ik}, \nno
d^\text{R}_u(k^2) &= \frac{i}{4k}\left(\log\frac{\mu_+/2 +ik}{\mu_+/2 -ik}-\log\frac{\mu_+/2 +i\eta k }{\mu_+/2 - i\eta k } \right),
\end{align}
for the $t$- and $u$-channel exchanges, respectively. The amplitude $f(k^2)$ can thus be written as 
\bea
\frac{1}{f(k^2)}= \frac{\bar{d}(k^2)-P(k^2)d^R(k^2)}{ n(k^2)}-ik = \frac{\tilde{d}(k^2)- g_0 d^R(k^2)}{\tilde n(k^2)+ g_0(L(k^2)-L_0)}-ik,
\eea
where $\tilde{n}(k^2)=g_0\frac{\bar{n}(k^2)}{P(k^2)}$ and $\tilde{d}(k^2)=g_0\frac{\bar{d}(k^2)}{P(k^2)}$ are rational functions with $g_0$ a parameter and $\tilde{n}(k^2=0)=1$. The rational functions can be expanded into polynomials and we call an $[m,n]$ approximant 
\begin{equation}
\label{eq:f}
\frac{1}{f_{[m,n]}(k^2)} = \frac{\sum_{i=0}^n\tilde{d}_ik^{2i}- g_0 d^\text{R}(k^2)}{1 +\sum_{j=1}^m\tilde{n}_j k^{2j} + g_0 (L(k^2)-L_0)}-ik,
\end{equation}
which can be regarded as an extension of ERE including the lhc. Here $\tilde d_i (i=0,\ldots,n)$ and $\tilde n_j (j=1,\ldots,m)$ are parameters to be determined from experimental or LQCD data.
In the following, we focus on the $[0,1]$ approximant, 
\bea\label{eq:f01}
f_{[0,1]}(k^2) = \dfrac{1}{\dfrac{\tilde d_0+\tilde d_1 k^2 -g_0 d^R(k^2)}{1+g_0(L(k^2)-L_0)}-ik},
\eea
which reduces to the traditional ERE when OPE is absent, i.e. $g_0=0$.

Using the ERE with the lhc in Eq.~\eqref{eq:f01}, we can derive expressions for the scattering length $a=f(k^2=0)$ and effective range $r=\dfrac{\td^2(1/f+ik)}{\td k^2}\Big|_{k=0}$. For the $t$-channel exchange:
\bea
a = \dfrac{1}{\tilde{d}_0+g_0/m_\text{ex}},\qquad r = 2\tilde{d}_1 - \dfrac{8g_0}{3m_\text{ex}^3}-\dfrac{4g_0}{m_\text{ex}^4 a},
\eea
and for the $u$-channel exchange:
\bea
a = \dfrac{1}{\tilde{d}_0+(1-\eta)g_0/\mu_+},\qquad r= 2\tilde{d}_1 -\dfrac{8 g_0}{3\mu_+^3}(1-\eta^3)-\dfrac{4g_0}{\mu_+^4 a}(1-\eta^4).
\eea

\subsection{Couplings to the exchanging particle}

Along the lhc, the discontinuity of the amplitude equals that of the tree-level OPE potential iterated in the LSE:
\bea
\im f = -\dfrac{\mu }{2\pi}\im T = -\dfrac{\mu }{2\pi}\im V_\text{OPE}.
\eea
The imaginary part of the OPE potential along the lhc is:
\bea
\im V_\text{OPE} (k^2) = -g_P\frac{\pi}{4k^2}\mathcal{F}_\ell,\qquad  k^2<k_\text{lhc}^2,
\eea
where $g_P=g_1g_2$ for $t$-channel exchange and $g_P=g^2$ for $u$-channel exchange. The factor $\mathcal{F}_\ell$ depends on the coupling structures: $\mathcal{F}_\ell = 1$ for $S$-wave vertices, while for $P$-wave vertices as in Eq.~\eqref{eq:L:P}, $\mathcal{F}_\ell= -\mex$ for $t$-channel and $\mathcal{F}_\ell= -\mu_\text{ex}^2$ for $u$-channel exchange.

While $d(k^2)$ contains only the rhc, $n(k^2)$ encodes the lhc and satisfies:
\bea
\im n(k^2) = -g_0\frac{\pi }{4k^2}, \qquad k ^2<k_\text{lhc}^2.
\eea
By matching the imaginary part of $f(k^2=k_\text{lhc}^2)$ to that of the potential $V_\text{OPE}$, we have \cite{Du:2024snq}
\bea\label{eq:gP}
g_P = -\frac{2\pi g_0}{\mu d^{0,\text{lhc}}\mathcal{F}_\ell},
\eea
where $d^{0,\text{lhc}}=d(k^2=k_\text{lhc}^2)$ is given by:
\bea
d^{0,\text{lhc}}=\tilde d_0-\frac{\tilde d_1 \mex}{4}+\frac{m_\text{ex}}{2}\left(1+\frac{g_0}{\mex}\right)+\frac{g_0 \log 2}{m_\text{ex}},
\eea 
for $t$-channel exchange, and:
\bea
d^{0,\text{lhc}}=\tilde d_0 -\frac{\tilde d_1 \mu_+^2}{4}+\frac{\mu_+}{2}\left( 1+\frac{g_0}{\mu_\text{ex}^2}\right) +\frac{g_0\log[2/(1+\eta)]}{\mu_+},
\eea
for $u$-channel exchange. Therefore, if we can determine the parameters in Eq.~\eqref{eq:f01} from experimental data or LQCD simulations, we can extract the coupling products $g_1g_2$ ($t$-channel) or $g^2$ ($u$-channel). Conversely, if these couplings are known from other processes, we can reduce the number of free parameters in $f_{[0,1]}(k^2)$ using Eq.~\eqref{eq:L:P}, which is particularly useful when data points are limited.

\subsection{Amplitude zeros}

Amplitude zeros~\cite{Castillejo:1955ed}, found in scattering processes involving three-body dynamics like in~\cite{Du:2023hlu}, may signal excited Efimov states~\cite{Braaten:2004rn} near threshold~\cite{Zhang:2023wdz}. In our parameterization, amplitude zeros occur when $n(k^2)=0$. For $t$-channel exchange in the $f_{[0,1]}$ approximant, a zero exists for $y>0$ at:
\bea
k_\text{zero}^2 = -\frac{\mex}{4}\left[1+\frac{1}{y}W(-e^{-y}y)\right],
\eea
where $y=1+\mex/g_0$ and $W$ is the Lambert function with two branches for real $y$.

For $g_0>0$ ($y>0$), we use the principal branch $W_0$.

For $g_0<-\mex$ ($0<y<1$), we use branch $W_{-1}$.

The zero can also be expressed in terms of observables:
\bea\label{eq:y}
y= 1 + \frac{1+\frac{4}{3}a m_\text{ex}(1-\log 4)-\frac{4\pi a \mex}{\mu g_P \mathcal{F}_\ell}   }{2+a m_\text{ex}\left(1-m_\text{ex}r/4\right)}. 
\eea

For $u$-channel exchange, the zero is determined by:
\bea
1+g_0\left[L_u(k_\text{zero}^2)+\frac{1}{\mu_\text{ex}^2}\right]=0.
\eea
When $|\Delta|\ll m_i$ (i.e., $\eta \ll 1$), the $t$-channel expressions apply to the $u$-channel case with $m_\text{ex}$ replaced by $\mu_+$.

\section{Application to the $DD^*$}\label{sec:DD}

As an illustrative example, we apply our parameterization $f_{[0,1]}$ to reproduce the $DD^*$ amplitudes obtained from LSE calculations in~\cite{Du:2023hlu}, where the pion and charmed meson masses are set to the LQCD values from~\cite{Padmanath:2022cvl}. The tetraquark state $T_{cc}^+(3875)$ was discovered by the LHCb Collaboration in 2021 through the $D^0D^0\pi^+$ invariant mass distribution~\cite{LHCb:2021vvq,LHCb:2021auc}. The $T_{cc}$ lies close to the $DD^*$ threshold, and approximately 90\% of the $D^0D^0\pi$ events contain a $D^{*+}$ state \cite{LHCb:2021auc}, making the $DD^*$ molecular interpretation particularly compelling.

In physical systems, the $D^{*+}$ ($D^{*0}$) decays into $D^0\pi^+/D^+\pi^0$ ($D^0\pi^0$), and the $T_{cc}^+$ is located above both the $D^0D^0\pi^+$ and $D^0D^+\pi^0$ thresholds. As demonstrated in~\cite{Du:2021zzh}, the one-pion exchange potential generates a three-body cut that significantly affects the precise determination of the $T_{cc}$ pole position. The $DD^*$ interaction has also been studied using LQCD, for example in~\cite{Padmanath:2022cvl,Chen:2022vpo,Lyu:2023xro}, where unphysical pion masses ($m_\pi>\Delta M=M_{D^*}-M_D$) are employed. The one-pion exchange potential generates an lhc that can substantially modify the pole structure \cite{Du:2023hlu}.

Using the mass parameters from~\cite{Padmanath:2022cvl,Du:2023hlu}, the $DD\pi$ three-body threshold is located at $\left({p_\text{3,rhc}}/{m_{DD^*}}\right)^2=0.019$, while the nearest lhc branch point occurs at $\left(p_\text{lhc}/{m_{DD^*}}\right)^2=-0.001$. This branch point lies very close to the $DD^*$ threshold $m_{DD^*}$ but above the lowest energy level. The extreme proximity of the lhc to the rhc significantly impacts the phase shift and consequently modifies the pole structure substantially \cite{Du:2023hlu}.

\begin{figure}[tb]
\centering
\includegraphics[width=0.48\linewidth]{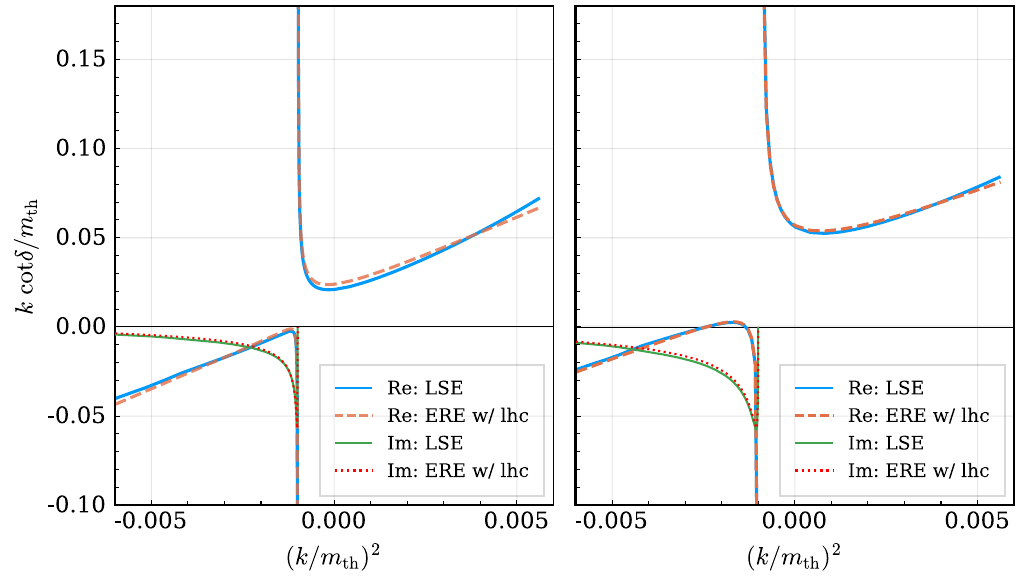}\includegraphics[width=0.52\linewidth]{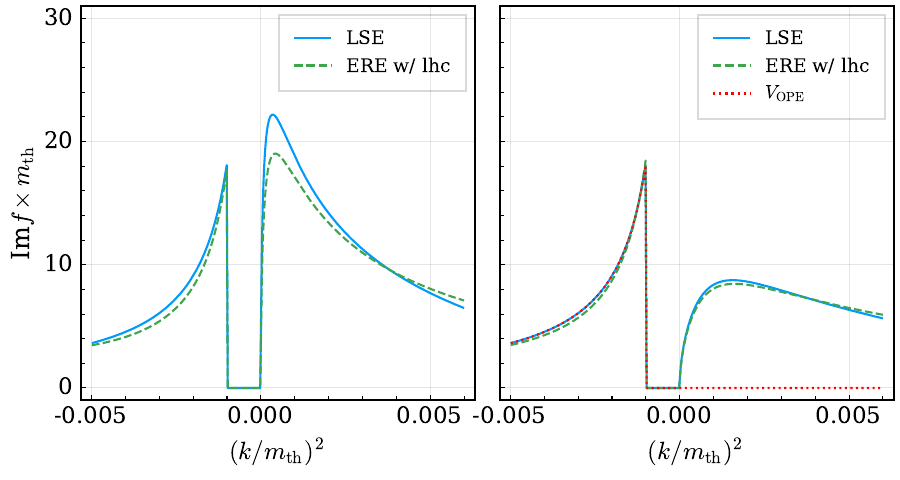}
\caption{Left: Comparison between our parameterization (Eq.~\eqref{eq:f01}, labeled as ``ERE w/ lhc") and LSE solutions from Ref.~\cite{Du:2023hlu}. The two sub-panels correspond to different fits from Fig.~3 in Ref.~\cite{Du:2023hlu}. 
    Solid curves show the real (blue) and imaginary (green) parts of $k\cot\delta$ from LSE calculations, while dashed and dotted curves represent the corresponding parts from our parameterization. Despite fitting only to real parts, the imaginary parts show remarkable agreement. Right: Comparison of $\im f$ between our parameterization and LSE calculations.}
\label{fig:im:VTf}
\end{figure}

In this work, we focus on reproducing the scattering amplitudes using Eq.~\eqref{eq:f01} instead of conducting a comprehensive analysis of experimental or LQCD data.
To determine the parameters $\tilde{d}_0$, $\tilde{d}_1$ and $g_0$, we fit the real part of the phase shift $k\cot\delta = 1/f+ik$ shown in Fig.~3 of~\cite{Du:2023hlu}.
We use 6 equidistant pseudo-data points for $(k/m_{DD^*})^2\in[-0.005,0.005]$ with uniform uncertainties.
The results are presented in Fig.~\ref{fig:im:VTf}, showing excellent agreement between our parameterization and the LSE solutions for both real and imaginary parts of the phase shift.
Our parameterization accurately captures key features including the amplitude zero (divergence of $k\cot\delta$) and the lhc.
The right panels of Fig.~\ref{fig:im:VTf} compare the imaginary parts of the amplitudes from both approaches.

Using the fitted parameters, we can extract the physical $P$-wave $DD^*\pi$ coupling $g$ through Eq.~\eqref{eq:gP}. For $u$-channel one-pion exchange, $g^2=g_{DD^*\pi}^2/(4F_\pi^2)$, where $g_{DD^*\pi}$ and $F_\pi$ are the $DD^*\pi$ coupling and pion decay constant from~\cite{Du:2023hlu}, respectively. Our two fits in Fig.~\ref{fig:im:VTf} yield $g^2=9.0~\text{GeV}^{-2}$ and $9.4~\text{GeV}^{-2}$, in excellent agreement with the input value of $9.2~\text{GeV}^{-2}$ used in the LSE calculations~\cite{Du:2023hlu}.

In summary, we have developed a model-independent parameterization for low-energy amplitudes that explicitly incorporates nonanalytic terms arising from OPE-induced lhc. Our approach naturally reduces to the traditional ERE in the absence of lhc, making it a proper generalization of the ERE formalism. This parameterization has wide-ranging applications and should prove particularly valuable in understanding near-threshold hadron resonances.

\end{document}